# SIMULTANEOUS QUALITATIVE AND QUANTITATIVE IDENTIFICATION OF MULTIPLE TARGETS BY POLYMERASE CHAIN REACTION ON MICROARRAY

Natalia V Zakharova[1,*], Alexei L Drobyshev[2,**], Dmitry A Khodakov[2], Anna I Myznikova[2], Irina A Gorban[3].

1 – State Research Institute of Genetics, 1st Dorozhny pr. 1, 117545 Moscow, Russian Federation
2 – Engelhardt Institute of Molecular Biology, Vavilov str. 32, 119991 Moscow, Russian Federation
3 – Moscow Institute of Physics and Technology, Institutsky per. 9, 141700 Dolgoprudny, Moscow region, Russian Federation
* - to whom correspondences should be addressed at NVZakharova@yandex.ru
** - to whom correspondences should be addressed at dro@newmail.ru

## ABSTRACT

An approach for multiplex qualitative and quantitative microarray-based PCR analysis has been proposed. The characteristics of PCR executed on a gel-based oligonucleotide microarray with immobilized forward primers and a single common reverse primer in solution were investigated for several DNA targets. One-stage multiplex on-chip PCR was studied for simultaneous amplification of herpes simplex viruses types 1 and 2, cytomegalovirus DNA, and bacteriophage lambda DNA as an internal control. Additionally the joint analysis of increased number of targets (with addition of *Chlamydia trachomatis*, *Mycoplasma hominis*, and *Ureaplasma urealyticum* DNA) was done in two-stage version of assay: first stage was in-tube PCR with target-specific primers, while the reverse ones contained 5'-adapter region; the second stage was on-chip amplification with immobilized target-specific forward primers and adapter as common reverse primer in solution. The possible application of one-stage reaction for human cDNA analysis was additionally demonstrated with utilization of a common poly-T-containing primer in solution. SYBR green I; and Cy-5 labeled dUTP were used for real-time and end-point detection of specific PCR products. The efficiencies of both one-stage and two-stage reactions was shown to be strongly dependent on magnesium and primers concentrations. Quantitative PCR in the both versions was studied with 10-fold serial dilutions of phage lambda DNA. The method enabled detection of 6 DNA copies per reaction for both versions of assay. The quantitative interval for one-stage reaction covered eight orders of concentration. The revealed significant effect of gel pad size on microarray PCR effectiveness has been discussed.

## INTRODUCTION

As a standard tool of modern molecular biology, PCR is routinely used in many microarray-related protocols, mainly in course of target preparation. It is aimed to increase the amount of analyzed target and decrease its excessive complexity. In the most simple scheme only one locus of target DNA is amplified and subsequently analyzed on microarray [1, 2]. For simultaneous analysis of several loci, the multiplex PCR with several pairs of primers could be used. However, in this case the degree of multiplexity is limited by potential primer dimmer formation. This effect is due to parasitic interactions in the pool of primers competing with primer-to-target interaction. It grows roughly like a square of number of primers and could be partially overcome by careful design of primer pairs [3]. Primer dimmer artifacts could be suppressed even further using various schemes with universal PCR.
Universal PCR is a multiplex polymerase chain reaction performed with a single or pair of universal primers to amplify a broad range of targets [4]. E. g., it could be designed as two-stage reaction, where the first stage occurs in the presence of chimerical primers consisting of locus-specific 3'-parts and universal sequences at their 5'-parts. Only universal parts are used as primers for the second stage amplification. One-stage reaction with chimerical primers could be performed with decreased annealing temperature at the first few cycles for amplification with specific 3'-parts of the primers, while at the subsequent cycles the temperature can be significantly increased enabling the annealing of only full-length chimerical primers [5]. These approaches selectively reduce the complexity of target by specific amplification of the loci of interest. The target complexity could be reduced in less specific manner without locus-specific sequences and without any risk of primer dimmer formation. In such an approach [6-8] genomic DNA is digested with restriction enzymes, common primer adapter is ligated and PCR is carried out with a single primer. As the restriction fragments longer than 1000 nucleotides are poorly amplified, the complexity of PCR product of digested DNA is lower than that of source DNA. Some modification of this procedure [9] can produce specific amplification of desired fragments. Also, whole-genomic DNA can be amplified without any reduction of complexity and the resultant product can be used e.g. for array-based comparative genomic hybridisation [10].

Universal PCR could be used for amplification of artificial sequences in whole genome genotyping experiments [11, 12]. These approaches imply the annealing and ligation of oligonucleotide probes at polymorphic sites of target DNA. The resultant probes contain sites complementary to universal primers as well as artificial sequences corresponding to particular site of polymorphism. Then the latter hybridize to complementary sequences of microarray upon the completion of PCR amplification of ligated probes with universal primers.

Solid phase PCR in microarray technology provides an elegant way to solution of many microarray-related problems but still has to be essentially elaborated for broad use in scientific community. In this approach at least some of immobilized oligonucleotides serve as primers in PCR carried out right over the microarray. Being immobilized on a solid phase, these primers cannot form dimmers regardless of their structure. For example single locus of target could be amplified with two liquid-phase primers while allele-specific solid-phase primers are elongated depending on target's sequence complementary to their 3´-end. The elongation of every solid-phase primer could be monitored either on-line [13] or upon the completion of reaction [14].

One of the more sophisticated versions of this approach implied the utilization of mineral oil to separate gel pads physically and to form thereby an array consisting of independent reaction volumes with immobilized primers [15]. This immobilization was reversible, i.e. 5´-immobilised solid phase primers containing 5´-triribonucleotide spacer were released to liquid phase by means of digestion of this spacer by ribonuclease A. Allele-specific primers were solid-phase-immobilized and internal in respect to ribonuclease-cleavable primers. The detection of allele-specific product was performed by means of post-PCR hybridization of fluorescently labeled oligonucleotides.

Another scheme with separation of reaction volumes implied an array of 3072 through-holes in 300μm thick stainless steel plate containing a set of gene-specific primer pairs for real time PCR-based expression profiling [16]. The higher parallelism (up to 300 000) was reached in the case of picoliter-scale plate manufactured by anisotropic etching of fiber optic face plate with resultant well density of 480 wells/mm² [17]. This microfabricated device enabled both liquid- and solid-phase PCR. In the latter case the solid-phase reverse primers are immobilized on Sepharose beads, while liquid phase contains both forward and reverse primers with 8-fold excess of forward ones. Later this technology was used (with substantial modifications) in novel method of massive parallel genome-wide DNA sequencing [18]. Nevertheless, the wide application of these arrays is still complicated due to some technical problems, particularly complication with their manufacturing.

Recently an original approach utilizing two-stage nested PCR and gel-based microarray for simultaneous quantitative analysis of three different viruses with specific pairs of primers was developed [19]. The approach was shown to be high-sensitive, specific, and its quantitative dynamic range covered up to six orders of concentrations. Nevertheless, the utilization of only specific primers essentially limits the number of targets for simultaneous analysis because of possible primer dimmer formation. In this respect, the present investigation is related to an improved approach, which enables quantitative analysis of increased number of nucleic targets by PCR on gel-based microarray with utilization of only one liquid-phase primer with common specificity. Direct one-stage reaction is studied for homologues viral DNAs and human cDNA. The two-stage indirect version of approach with introduction of universal nucleotide sequence via first-stage PCR with chimerical primers is developed for non-homologues targets. The detailed optimization of PCR conditions performed in the frame of the investigation has revealed several characteristics significant for high-effective amplification on gel-based microarray.

**MATERIALS AND METHODS**

**Microarrays**

Oligonucleotide microarrays were fabricated by spotting of gel polymerization solutions [20] containing different solid-phase primers onto glass slides (Corning 2947 Micro Slides, Corning, USA) using QArray spotter (Genetix, England). Glass slides treatment, gel polymerization and washing conditions were described earlier [19]. The fabricated microarrays consisted of spatially separated gel pads with varied 5`-immobilised forward primers specific to different DNA targets. In the most of experiments gel pads with diameter of 600 μm were used (unless otherwise specified). Optimal primer concentrations have been described in "Results".

**Primers**

Primers were designed using Oligo 6 (Molecular Biology Insights, USA) and DNA sequences available from public databases (Table 1). Primers were synthesized on an ABI-394 DNA/RNA synthesizer (Applied Biosystems, USA) by standard phosphoramidite method and purified by reverse phase high performance liquid chromatography on C18-Nucleosil (5 μm, 4.6 x 250 mm) columns (Sigma, USA). 5'-Amino-Modifier C6 (Glen Research, USA) was used for introduction of amino group into 5'-nucleotides of primers for on-chip immobilization (below referred as "solid-phase primers"). Primers without such amino groups are referred below as "liquid-phase primers".

The solid-phase forward primers were designed to contain highly specific oligonucleotide sequences free from any essential internal hairpin structures, the $T_m$ of the primers were in the range of 55-60 °C (18-23 nucleotides). The varying nucleotides in allele-specific primers designed for HSV-1 and HSV-2 were located on 3`-ends.

Two sets of unmodified liquid-phase primers were designed. A set for one-stage on-chip PCR of homologues targets (Table 1) consisted of a common reverse primer specific to HSV-1, HSV-2, and CMV and a separate reverse primer specific to lambda DNA (used as an internal control). Additionally, solid-phase primers for human β-actin and hypoxanthine phosphoribosyltransferase (Actb_F_I and Hprt_F_I, respectively, Table 1) were designed to be used in combination with common reverse liquid-phase primer T_R. Another set was designed for joint multiplex two-stage PCR of homologues HSV-1, HSV-2, CMV, and non-homologues *U. urealyticum*, *M. hominis*, *C. trachomatis*, and bacteriophage lambda DNAs (Table 1). Target-specific forward primers designed for in-tube first-stage reaction were outer relative to second-stage solid-phase forward ones. First-stage reverse primers were chimerical and contained 3´- target-specific and 5´- universal adapter sequences in order to apply the common reverse primer for further second-stage on-chip PCR (Table 1).

**Target Nucleic Acids**

HSV-1, HSV-2, CMV, *U. urealyticum*, *M. hominis*, and *C. trachomatis* DNA preparations isolated from anonymous urea, cervical, and blood clinical samples were kindly provided by Scientific Center of Obstetrics, Gynecology and Perinatology, Russian Academy of Medical Sciences. Three samples of each target were utilized to validate the performance of the assay. Bacteriophage lambda DNA (Fermentas, Lithuania) with initial concentration of $6\times10^{12}$ copy per 1 mL was used in optimization experiments and quantitative PCR studies. The 10-fold serial dilutions of lambda DNA was used to obtain concentrations from $6\times10^{8}$ to $6\times10^{0}$ copy per assay. Total RNA was isolated from K562 myelogenous leukemia cells with RNeasy Mini Kit (Qiagen, Germany) according to the manufacturer's protocol.

**Reverse Transcription**

The reaction mixture for reverse transcription (RT) of total RNA contained 1× StrataScript buffer (Stratagene, USA), 0.2 mM each dATP, dCTP, and dGTP, 0.6 mM dUTP, 2 μM RT-primer (T_R, Table 1), 10 units StrataScript® Reverse Transcriptase, 20 units RNasin® and 10 μL of total RNA preparation per 10 μL reaction. The reverse transcription was performed at 42 °C for 90 min with final incubation at 95 °C for 10 min. One μL of 4-fold diluted cDNA preparation was used for further on-chip PCR.

**PCR on Microarray**

One-step PCR was performed inside a microarray chamber formed by sealing the microchip slide with a cover glass slide by a Frame-Seal (Bio-Rad, USA). The reaction mixture consisted of 70 mM Tris-HCl (pH 8.6), 16.6 mM $(NH_4)_2SO_4$, 1.5 mM $MgCl_2$, 0.2 mM each dATP, dCTP, and dGTP, 0.6 mM dUTP (Sileks, Russia), 0.2 μM each solution-phase reverse primer (Table 1), 0.01% acetylated BSA (New England BioLabs, USA), 0.5 unit Platinum® *Taq* DNA Polymerase (Invitrogen, USA), 10 μL of template DNA, and 1× SYBR® green I (Molecular Probes, USA) or 2 μM Cy5-dUTP per 25 μl reaction. After initial incubation at 95 °C for 2 min, 60 cycles (by default, 40 s at 95 °C, 20 s at 55 °C and 45 s at 72 °C) were performed.

For two-stage semi-nested PCR the first stage of in-tube amplification with outer forward and chimerical reverse primers was performed with reaction mixture containing 70 mM Tris-HCl (pH 8.6), 16.6 mM $(NH_4)_2SO_4$, 2.5 mM $MgCl_2$, 0.2 mM each dATP, dCTP, and dGTP, 0.6 mM dUTP (Sileks, Russia), 0.02 μM each primer (Table 1), 0.5 units Platinum® *Taq* DNA Polymerase (Invitrogen, USA), and 10 μL of template DNA per 25 μl reaction. PCR was performed as follows: initial incubation for 2 min at 95 °C; 10 three-step cycles with denaturing for 20 s at 95 °C, annealing for 30 s at 55 °C, and elongation for 30 s at 72 °C; 25 three-step cycles with annealing at 68 °C and the same other conditions; and final incubation at 72 °C for 5 min. The second stage was performed on-chip in the conditions described above for one-step reaction. One μL of the first-stage PCR product was used as a template.

**Data Acquisition and Processing**

For real-time measurements of SYBR® green I fluorescence an original experimental setup consisted of fluorescent microscope equipped with thermotable, a CCD camera, a temperature controller [13] and specialized software ImageExpress® (Biochip-EIMB, Russia) was used. Images were acquired every cycle at the end of elongation step. The melting curves were obtained by measuring the fluorescence with 10 second interval upon the gradual increasing of temperature from 72 °C to 95 °C at the rate of 3 °C/min. The raw fluorescence data ($F$) were normalized according to formula $(F-F_{min})/(F_{max}-F_{min})$.

The end-point SYBR® green I and Cy5 fluorescence were measured with specialized software ImageWare® (Biochip-EIMB, Russia). SYBR® green I fluorescence was detected without removing of reaction mixture from microarray at 72 °C to avoid the contribution of non-specific hybridization. To measure Cy5 fluorescence slides were unsealed, and unincorporated Cy5-dUTPs were removed by electrophoresis [21] in TAE-buffer at 15 V/cm for 5 min.

## RESULTS

**Design of the Assay**

The scheme of PCR on a microarray with immobilized specific forward primers and a common liquid-phase reverse one is illustrated on Fig. 1. Target-specific forward primers are immobilized in gel pads of microarray. This type of immobilization provides every element of microarray with primers in amount sufficient for PCR reaction [13, 19]. In this study we use a single universal primer matching all the targets under investigation (Fig 1A). This combination with spatially separated target-specific solid-phase primers enables PCR with (theoretically) unlimited multiplexity as immobilized solid-phase primers cannot form primer dimmers. In course of PCR, target DNA anneals to immobilized forward primers enabling their elongation by polymerase (Fig. 1B). These elongated solid-phase anchored primers serve as templates for complementary strands synthesis by means of elongation of liquid-phase reverse primers (Fig 1C). The accumulated double stranded solid phase anchored product could be visualized by DNA stain (SYBR® green I) or by incorporation of fluorescently labeled dNTRs (not shown).

Such an approach seems to be a powerful tool for parallel analysis of homologous sequences, for example allele genes and gene mutations. The method could be also useful for analysis of targets possessing some common sequences, like poly-A tail in eukaryotic mRNA. The utilization of chimerical primers for an introduction of a universal adapter oligonucleotide sequence by PCR (Fig. 2) or combined restriction-ligation reactions gives another opportunity to apply on-chip PCR for analysis of complex mixtures of non-homologous nucleic targets. In this connection, the present study was focused on estimations of two on-chip PCR approaches: (i) direct one-stage PCR with one primer specific to several targets, and (ii) indirect two-stage reaction with introduction of universal sequence via in-tube PCR with chimerical primers (Fig 2).

The set of six pre-natal/new-born pathogen DNAs (HSV-1, HSV-2, CMV, *U. urealyticum*, *M. hominis*, and *C. trachomatis*) and bacteriophage lambda DNA (included for imitation of internal control) was considered as a model system since it contain both homologous and non-homologous targets. Furthermore, the detailed studies on some on-chip PCR characteristics required several comparative experiments with different nucleic targets (e.g. eukaryotic cDNA) and individually designed microarrays (see bellow).

Two different dyes were investigated for identification of specific on-chip PCR products. Non-specific dsDNA stain SYBR® green I was suitable for both real-time and end-point detection schemes as optimal measurements condition (72 °C) provided melting of non-specific short-chain structures and enabled the reliable detection of stable specific long-chain DNA products. The real-time monitoring of PCR by means of Cy5-dUTP was technically difficult because of strong fluorescence of unincorporated Cy5-dUTP and essential decrease of Cy5 fluorescence under PCR temperature conditions. Nevertheless, Cy5-dUTP was convenient for end-point detection of accumulated Cy5-labeled specific products covalently attached to microarray. The electrophoretical removal of unincorporated Cy5-dUTP from microarrays after the PCR enables improved discrimination between positive and negative gel pads. Figure 3 A and B represents the end-point results obtained with the same DNA targets in two parallel on-chip experiments performed with Cy5-dUTP and SYBR® green I, respectively.

**Optimization of On-chip PCR**

One of significant disadvantage of solid-phase PCR systems described so far is their low efficiency in comparison with traditional in-tube PCR [13, 22-24]. In order to improve this characteristic several factors were varied in this study. The use of SYBR® green I for real-time monitoring of on-chip PCR enables the direct estimations of effects of different parameters on PCR (Fig. 4). No significant change was found when 70 mM Tris-HCl (pH 8.6) and 16.6 mM $(NH_4)_2SO_4$ PCR buffer was replaced by 10 mM Tris-HCl (pH 8.6) and 100 mM KCl one, and dNTPs concentrations were varied in the range from 0.1 to 0.4 mM for dATP, dCTP, and dGTP (0.3-1.2 mM dUTP, respectively) (data not shown). The concentration of liquid-phase primers strongly affected the threshold cycle number although did not significantly change the amplification coefficient (Fig. 4A, all the curves have the same slope). However the use of solid-phase primers in concentrations less than 100 μM essentially influences the both characteristics (Fig. 4B).

The simultaneous presence of more then one liquid-phase primer in reaction mixture significantly increases the chance of primer dimmers formation and leads to accumulation of non-specific double stranded PCR products competing with specific ones for SYBR® green I binding. This possibly caused the decrease of specific solid-phase PCR signals and accordant increase of liquid-phase fluorescence around the gel pad (Fig. 4C, solid and opened circles, respectively) observed after 30-35th cycle for the reaction with two (HPR-R and BL-R; Table 1) liquid-phase primers and $10^7$ copies of phage lambda DNA. The effect was not observed when only one liquid-phase primer (BL-R, Table 1) was used (Fig. 4C, squares). The addition of PCR enhancer solution to reaction mixture with two primers eliminated this effect as well, but some increase of threshold cycle number was observed (Fig 4C, triangles).

Magnesium concentration was varied from 1mM to 4mM and 1.5mM was found to be optimal in terms of maximal fluorescent signal (Fig. 4D), while the threshold cycle was not significantly affected (not shown). Still the optimal concentration could be various for different sets of simultaneously amplified targets and must be adjusted individually. Interestingly that the optimal magnesium concentration obtained for on-chip reaction was not optimal for in-tube reaction performed in identical conditions (1× SYBR® green I) with the same DNA target (phage lambda DNA). The known inhibitory effect of SYBR® green I [25] requires the increase of magnesium concentration up to 4 mM in case of in-tube reaction only. This discrepancy for on-chip and in-tube reactions could imply some anti-inhibitory properties of gel pads. Nevertheless, SYBR® green I in concentrations higher than one fold inhibits on-chip reaction as well (data not shown).

**Quantitative On-Chip PCR**

Figure 5A demonstrates the typical PCR curves obtained with SYBR® green I in one-stage on-chip amplification of 10-fold dilution series (from 6 to $6\times10^8$ copies) of phage lambda DNA. The threshold cycle ($C_T$) values were determined from the interceptions of the curves with the threshold line (corresponded to 1/20 of maximal increase of fluorescence) and were used to obtain the calibration curve (Fig. 5B) for further quantitative determinations of initial target amount. The calibration curve for the two-stage on-chip PCR was generated in the same way with first-stage DNA products as templates.

The standard deviations of experimental data from linear fit ($C_T$ versus logarithm of input copy number) were 0.85 and 0.37 cycles (with maximal deviations of 1.32 and 0.6 cycles) for one-stage and two-stage experiments, respectively. Both calibration curves have virtually the same slope (3.32 and 3.36 cycles per decade of input copy number, Fig 5B, solid and open circles, respectively) but the two stages scheme provides relatively narrow dynamic range (4 orders of magnitude) as accurate $C_T$ measurements were technically difficult for concentrations of $\geq10^5$ copies per reaction after 35 amplification cycles performed during the first (in-tube) stage. The dynamic range could be broadened by decreasing the number of in-tube stage cycle amount [19]. Maximally it tends to 8 orders of magnitude, like in the case of direct one-stage on-chip PCR (i.e. 0 cycles of preliminary amplification, Fig. 5B). The certain inhibition of PCR was observed for $6\times10^9$ copies (data not shown), while the precise detection limit value was not detected in the frame of the investigation.

**Multiplex On-Chip PCR**

Figure 6 shows the end-point results of five individual on-chip experiments on detection of several simultaneously amplified DNA targets. Bacteriophage lambda DNA was added to all the samples analyzed to imitate the internal control. For higher multiplexity two of the shown experiments were performed with artificially prepared mixtures of different DNA targets (Fig. 6 B and C). The results shown were obtained both with SYBR® green I (Fig. 6 B and C) and Cy5 labelled dUTP (Fig. 6 D-F), in one-stage (Fig. 6 B and D) and second-stage (Fig. 6 C, E, and D) on-chip PCR. Totally three samples of each target were utilized to validate the performance of the assay. The amplification of all analyzed samples was highly specific both in one-stage and two-stage reactions. This was especially important for relative *Herpesviridae* targets, which were additionally verified by co-amplification in all possible combinations (Fig. 6 B). No false-positive results were obtained with non-specific DNAs utilized in control experiments.

To estimate the possible competitions between simultaneous PCR of different DNA targets, the results of individual amplification of lambda DNA were compared with results of multiplex reactions. Figure 7 illustrates the reasonable concordance of PCR results obtained for series of lambda DNA concentrations (from 6 to $6\times10^5$ copies) in the absence or upon simultaneous amplification of abundant HSV-2 DNA ($10^6$ copies per reaction). These data as well as similar results obtained with several sets of three simultaneously amplified specific targets (data not shown) imply no essential competition between accumulations of individual PCR products upon joint multiplex analysis.

**On-Chip PCR with Eukaryotic cDNA and Gel Pad Size Impact**

Eukaryotic cDNA is another suitable object for application of the approach proposed, and its one-stage format, in particular. Poly-A tail provides a natural site for common reverse primer binding, semi-quantitative results of real time solid-phase PCR on microarray could provide biologically valuable information on abundance of particular mRNA species. However, in comparison with applications described above, this one has some additional challenges.

The first challenge is related to low melting temperatures of poly-A structures. This complicates the selection of gene-specific forward primers as they should have the same melting temperature as poly-T reverse primer. For this reason the reverse transcription in this investigation was carried out with the primer consisting of 3´-$T_{25}$ part and 5´- T7 promoter part (primer T_R, Table 1). This primer was also used as a reverse liquid-phase in a subsequent one-stage on-chip PCR reaction.

Second point relates to high complexity of eukaryotic cDNA. It is a pool of thousands of individual cDNA species possessing common liquid-phase reverse primer annealing site. Only tiny share of individual cDNA in this pool are specific to solid-phase forward primers. This circumstance challenges the very possibility of successful solid phase on-chip PCR.

Third challenge relates to high number of potentially interesting genes. In long term, the expression profiling could require a high density microarray with thousands of elements. In this case the decrease of gel pads physical size could be advantageous.

In order to address these challenges, microarray of gel spots with diameters of 600μm, 450μm, and 250μm containing immobilized forward primers specific to *Actb* and *Hprt* genes was manufactured and investigated in one stage on-chip PCR. Sixty cycles of PCR were carried out with T_R primer (Table 1) and cDNA from K562 cells as follows: 20s at 95°C, 40s at 54°C, 45s at 68°C. Microarray fluorescent images were taken at the end of elongation stage. PCR curves revealed a good sigmoid shape (not shown) with threshold cycles indicated at Table 2 (average from 3 experiments). Threshold cycles were found to be strongly dependent on the size of spot. For 250μm spots $C_T$ was unacceptably high hampering accurate assessment of low abundance genes. For 600μm spots $C_T$ was the highest, and the discrimination between high and low abundance genes was decreased (26 and 28 cycles for *Actb* and *Hprt*, respectively). Post-PCR melting curves revealed sharp transition with melting temperature 87°C and 90°C for *Actb* and *Hprt*, respectively.

**DISCUSSION**

Microarray platforms have been widely investigated as a tool for performing of complicated multiplex analyses of nucleic acids [11]. Several microarray-based PCR approaches developed so far provide just qualitative methods which often suffer from low efficiency in comparison with traditional in-tube PCR [13-15, 22-24, 26-30]. Thereupon, the present investigation was focused on a study of several characteristics of PCR on a microarray of gel-based elements in order to develop the improved approach for multiplex DNA analysis. This chapter considers the application of microarray based solid phase PCR with a common liquid-phase primer to several relevant fields of modern life science, like detection of microorganisms, DNA quantification, eukaryotic expression profiling.

Previously the effective PCR on a microarray of gel elements was shown to occur only in the presence of both pairwise primers in liquid phase [13]. But for "two primers per target" scheme the probability of primer dimmer formation increases rapidly with the number of simultaneously analyzed targets. The recently developed two-stage nested PCR approach for simultaneous quantitative analysis of three blood-born viruses [19] also utilized only specific pairs of primers for each nucleic target: the first-stage reaction with outer primers was performed in tube, the second stage was carried out on-chip with immobilized inner forward primers and liquid-phase inner reverse ones.

In contrast to previous studies [19], here we detect and quantify DNA of several microorganisms in a single stage format of PCR (Fig. 3A, Fig. 6B and D, and Fig. 5B, solid circles, respectively). This format minimizes the risk of contamination and maximizes the dynamic range of assay. The key advantage of the investigation is utilization of only one universal liquid-phase primer, what (theoretically) enables the significant increase of amount of nucleic targets for simultaneous analysis. This goal was achieved in two ways: (i) directly by design of the primer possessing the common specificity to homologues DNAs, and (ii) indirectly by introduction of universal sequence into non-homologues targets via PCR with chimerical primers. The second way represents the version of semi-nested PCR format, which is still advantageous in comparison with approaches described for the gel element microarray earlier [13, 19] since it enables the analysis of increased number of targets. At the same time two-stage PCR provides more accurate results, as it inherent in conventional two-stage in-tube format. The effectiveness the on-chip PCR under conditions optimized in the frame of the investigation is shown to be very high and close to conventional in tube methods one.

One of the intriguing results of the study is the demonstration of the possibility of performing of microarray based PCR with eukaryotic cDNA. However, the effectiveness of the reaction ($C_T$ value) was found to be strongly dependent on the gel pad size. To understand this phenomenon the difference between conventional liquid-phase (in tube) and microarray-based solid-phase PCR should be considered. Conventional liquid-phase PCR includes two pairwise primers uniformly distributed across the reaction volume. While in the case of microarray-based PCR the only one of the primers is distributed in liquid phase, and the opposite one is immobilized within the gel pad, which is just a tiny fraction of total reaction volume. This circumstance complicates the interaction of target DNA strands with complementary solid-phase primers. After the elongation step, target DNA copies diffuse out of gel pad (Fig. 8A and B). To serve as templates again they have to diffuse back into the gel pad to anneal with forward primer during the next annealing stage (Fig. 8C). Liquid-phase PCR is free from this limitation as every copy of the target could serve as a template every next cycle. In the case of gel-based microarray PCR this effect must be more expressive for small sized gel pads. To demonstrate it a following simple model of gel pad with hybridized PCR product after elongation stage can be considered (Fig. 8A). If template DNA at the melting stage diffuse out of gel pad to some effective area (located at range $D$ from the surface of gel pad, Fig. 8B) and only molecules from the absorbance area (located at range $a$ from the surface of gel pad, Fig 8C) are able to diffuse back upon the annealing conclusions, then the portion $p$ of PCR product available as template at the next cycle could be estimated as the ratio of absorbance and diffusion areas:

$$p=(a+r)^2/(D+r)^2,$$

provided $(a+r)>>h$ and $(D+r)>>h$, where $h=250\mu m$ is the thickness of solution layer over the micoarray, and $r$ is the pad radius. Obviously, the higher $r$ gives the higher $p$ causing better yield of PCR cycle and lower $C_T$ value, in accordance with data from Table 2. Also, the higher concentration of immobilized forward primer should result in better absorbance ability of a gel pad and lower $C_T$ (in accordance with results of Fig. 4B). The lower yield of PCR cycle could be the reason for higher discrimination between $C_T$ value determined for *Actb* and *Hprt* genes in the case of small (450μm and 250μm) gel pads (Table 2). The observed discrimination for 600μm pads ($\Delta C_T=28-26=2$) is close to the data obtained earlier with Taqman PCR for K562 cell line ([31]: $\Delta C_T=2.7$ with $C_T=18.9$ and 21.6 for *Actb* and *Hprt*, respectively). Nevertheless, the smaller sized pads could be preferable if the higher $\Delta C_T$ discrimination is desired for some kind of reasons.

**CONCLUSION**

This chapter presents a new powerful approach applying the gel element based microarrays for high-effective multiplex PCR with utilization of multiple solid-phase immobilized primers and a common universal liquid-phase one. The results obtained demonstrate high sensitivity (<10 DNA copies per reaction) and specificity of the method. The approach enables the reliable qualitative and quantitative determinations of multiple nucleic acid targets analyzed simultaneously. The quantitative dynamic range covers up to eight orders of concentrations. The approach is shown to be suitable for direct application for parallel analysis of relevant genes possessing homologous sequences (as exemplified with DNA polymerase (*pol*) gene of three *Herpesviridae* viruses). Furthermore, non-homologous sequences can be analyzed by the microarray PCR indirectly if two-stage format with introduction of a universal sequence is applied (as shown for the set consisting of HSV-1, HSV-2, CMV, *U. urealyticum, M. hominis, C. trachomatis,* and bacteriophage lambda DNAs). The direct eukaryotic mRNA expression profiling has been considered as one of the advantageous fields for application of PCR on a microarray and requires further particular investigation.

**ACKNOWLEDGMENTS**

We are grateful to members of Scientific Center of Obstetrics, Gynecology and Perinatology, Russian Academy of Medical Sciences for providing DNA preparations isolated from human clinical samples, and Dr. Svetlana Strunina for preparation of total RNA from K562 cells. We also thank EIMB RAS colleagues for their technical support.

**Table 1. The sequences of primers used in the assay.**

| Primer[a] | Sequence from 5´ to 3´ | Position[b] |
|---|---|---|
| **Solid-phase forward primers** | | |
| HSV-1_F_I | CGG ACT CCA TAT TTG TGC | 3240 |
| HSV-2_F_I | CGG ACT CCA TTT TCG TTT | 2675 |
| CMV_F_I | CGT GTT TGT CCG CTT TCG | 2739 |
| UU_F_I | ACA AAC CCA ACT ATT CCA TAT AC | 1817 |
| MH_F_I | CTA TTC TAG AAA GCC ACG ATG C | 140 |
| CT_F_I | GTT AAT ACC CGC TGG ATT TGA G | 467 |
| BL_F_I | AGG AGC TGG ACT TTA CTG ATG | 1561 |
| Actb_F_I | TAT TTT GAA TGA TGA GCC TTC GTG | 1675 |
| Hprt_F_I | TAA GAA GTT TTG TTC TGT CCT GG | 1189 |
| **One-step solution-phase reverse primers** | | |
| HRP-R | CGC ACC AGA TCC ACG CCC TT | 3456 (HSV-1) |
| | | 2891 (HSV-2) |
| | | 2951 (CMV) |
| BL-R | GTC AGG TGG CTC AAT CTC TTC | 1670 |
| T_R | GGC CAG TGA ATT GTA ATA CGA CTC ACT ATA GGG AGG CGG (T)$_{25}$ | |
| **Two-stage PCR primers** | | |
| HRP_OUT_F[c] | GCG GGT CAT CTA CGG GGA CAC GGA C | 2712 (CMV) |
| | | 3220 (HSV-1) |
| | | 2655 (HSV-2) |
| HSV_u_R | ***AGG TGG TGA ACG GGC TGT A**CG GTT GAT AAA CGC GCA GTT G* | 3483 (HSV-1) |
| | | 2918 (HSV-2) |
| CMV_u_R | ***AGG TGG TGA ACG GGC TGT A**GC CGA TGT AAC GTT TCT TGC* | 2901 |
| UU_OUT_F | CAG AAG GTG CTG GTG GTG GAC A | 1743 |
| UU_u_R | ***AGG TGG TGA ACG GGC TGT A**CT TCT GGA ACC TTA GGA TTT AAG TG* | 1910 |
| MH_OUT_F | GAC TAC ATT ACA CCA GCT CGT TTA GAC | 97 |
| MH_u_R | ***AGG TGG TGA ACG GGC TGT A**CA ACA ACG TTG ATT CCT CTG T* | 220 |
| CT_OUT_F | GCG TGT GTG ATG AAG CTC TAG GGT TG | 407 |
| CT_u_R | ***AGG TGG TGA ACG GGC TGT A**AA CTA ACT TAC CTT TCC GCC TAC* | 608 |
| BL_OUT_F | ATT GAG CGT GCA GCC AGT GA | 1377 |
| BL_u_R | AGG TGG TGA ACG GGC TGT ACG CTG T | 1716 |
| R_u | AGG TGG TGA ACG GGC TGT A | |

[a] F, forward; R, reverse; I, immobilized; OUT, first stage PCR primers (outers); u, universal sequence; HSV-1, HSV-2, herpesviruses type 1 and 2, respectively; CMV, cytomegalovirus; UU, *U. urealyticum*; MH, *M. hominis*; CT, *C. trachomatis*; BL, bacteriophag lambda; Actb, human β-actin gene; Hprt, human hypoxanthine phosphoribosyltransferase gene.

[b] The figure indicates the 5´ terminus nucleotide position number according to Gen Bank accession no. M10792 (HSV-1, *pol* gene), AY038367 (HSV-2, *pol* gene), DQ180391 (CMV, *pol* gene), AF085728 (*U. urealyticum*, *ureA* gene), D13314 (*M. hominis*, *adi* gene), DQ019310 (*C. trachomatis*, 16S rRNA gene), J02459 (enterobacteria phage lambda, complete genome), Ensemble Transcript ENST00000331789 (human β-actin gene), and ENST00000298556 (human hypoxanthine phosphoribosyltransferase gene).

[c] The utilized common first stage forward *Herpesviridae* primer corresponds to CMV sequence while analogous HSV sequences contain some varied nucleotides close to 5´ terminus.

**Table 2. Threshold cycle numbers for different spot diameters in human cDNA experiments**

| Spot diameter, μm | $C_T$ for *Actb* | $C_T$ for *Hprt* |
|---|---|---|
| 600 | 26 | 28 |
| 450 | 30 | 40 |
| 250 | 39 | 50 |

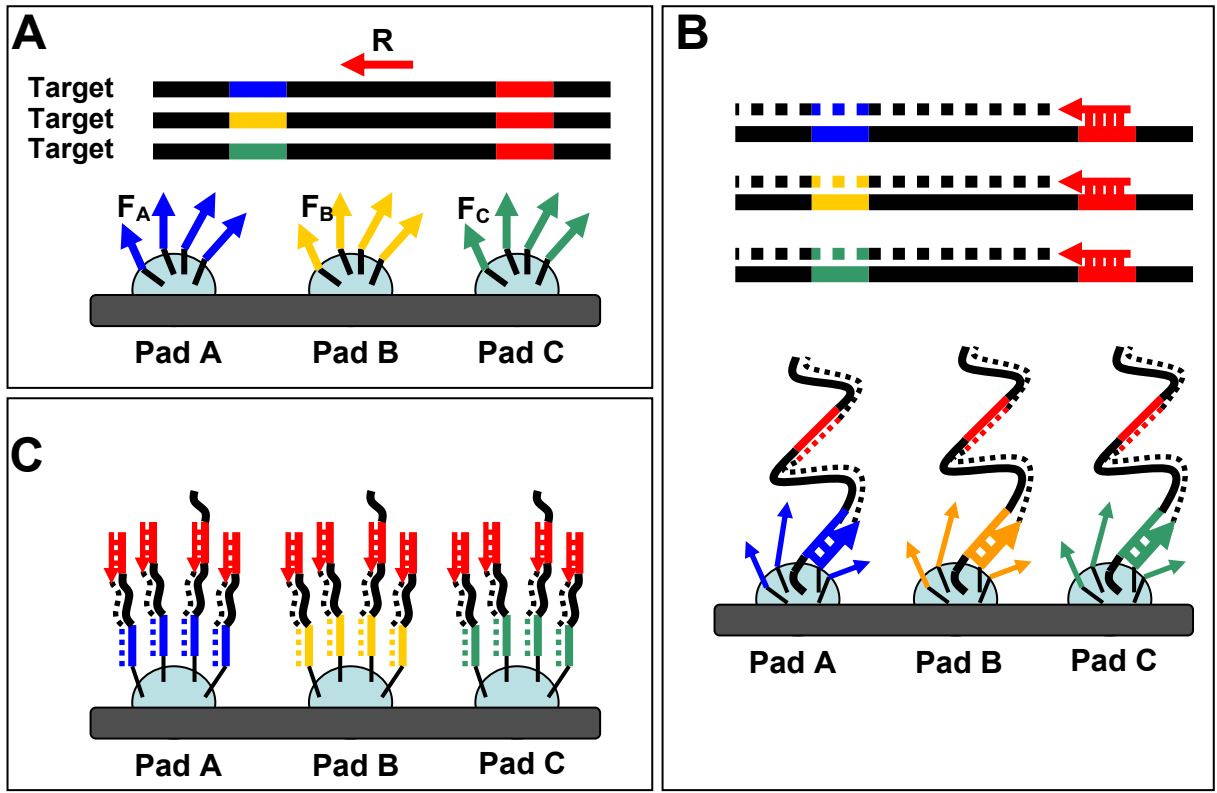

**Fig. 1. Scheme of PCR on gel pad oligonucleotide microarray.**
(A) A set of homologues DNA targets and array of gel-immobilised solid-phase forward primers for their identification. Every target contains the sequence specific to immobilized forward primer (both depicted with the same colour). Liquid phase contains only one reverse primer (red coloured) specific to all the targets in pool. (B) Target DNAs anneal to specific solid-phase forward primers enabling their elongation and formation of a complementary second strand. These copies serve as templates for subsequent amplification. (C) Elongated solid-phase anchored strands serve as templates for liquid-phase reverse primers complementary elongation.

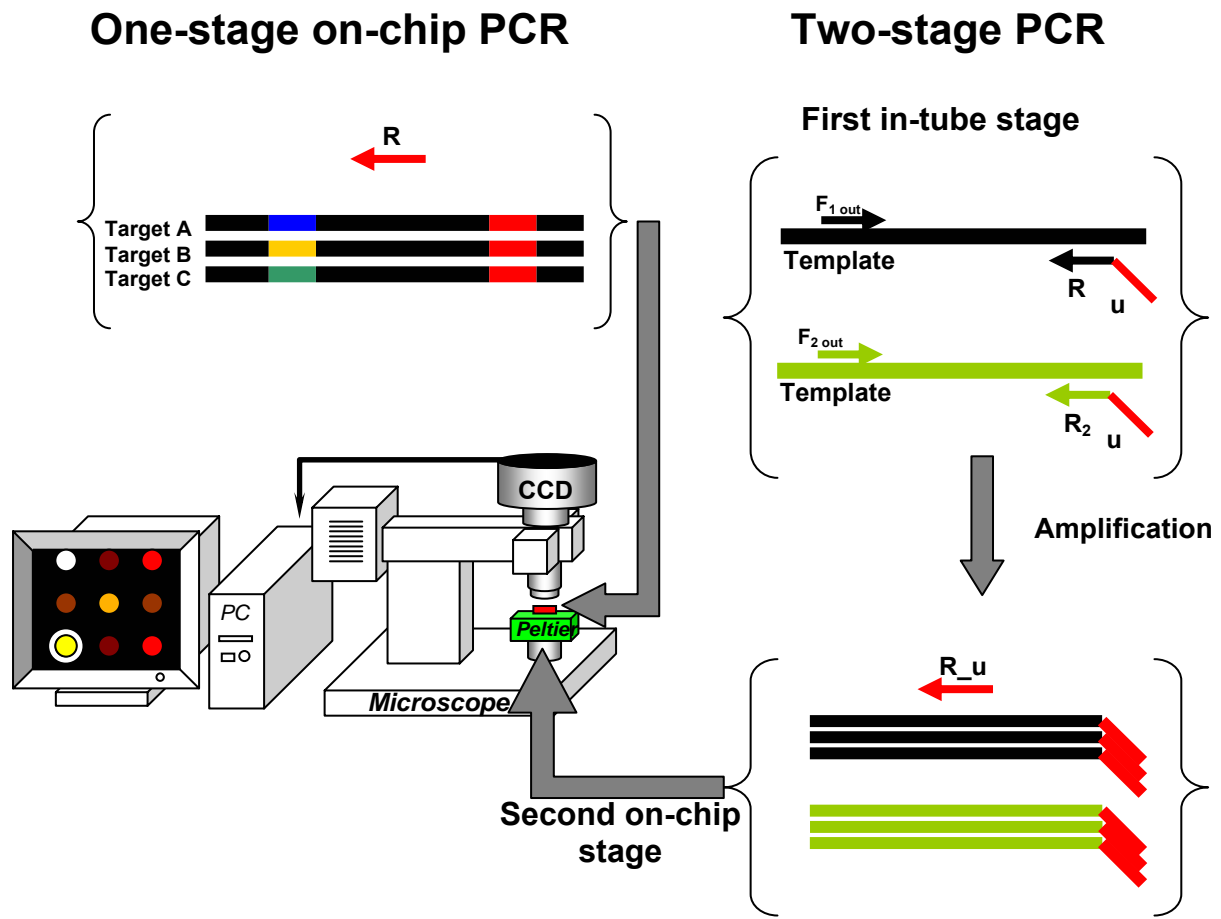

**Fig. 2. The scheme of one-stage and two-stage analysis via on-chip PCR.**
Homologues DNA targets can be analysed directly by one-stage on-chip PCR (left) using single liquid-phase primer (R) with common specificity. Non-homologues targets analysis requires two stages PCR (right). First in-tube stage of PCR is necessary for introduction of a universal sequence (u) into DNA products amplified from specific targets. The second on-chip stage of PCR can be performed with a common liquid-phase primer (R_u).

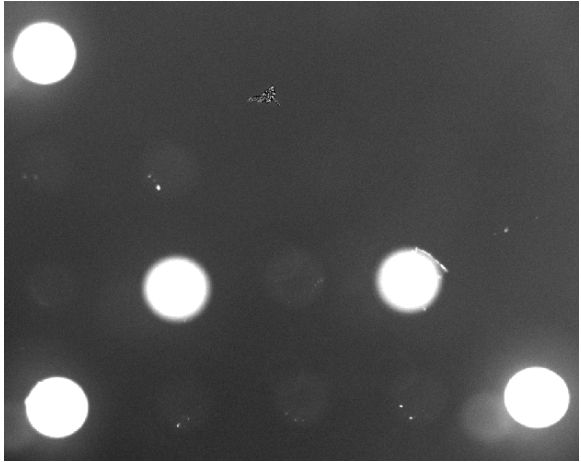 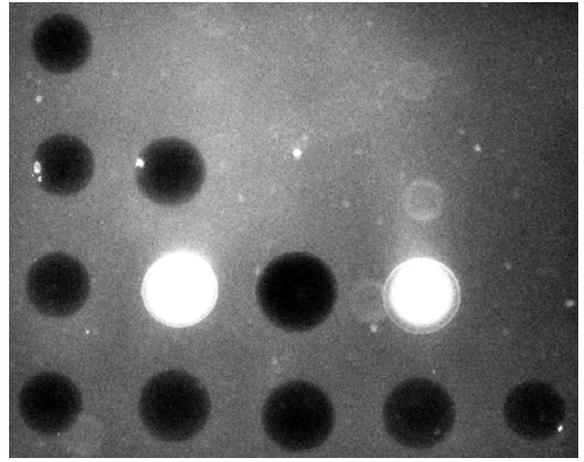

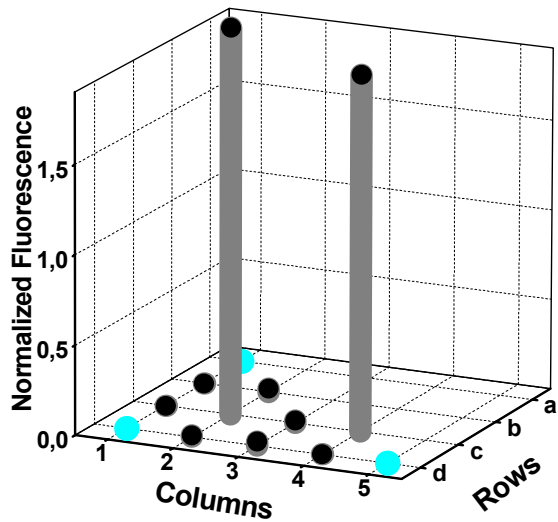 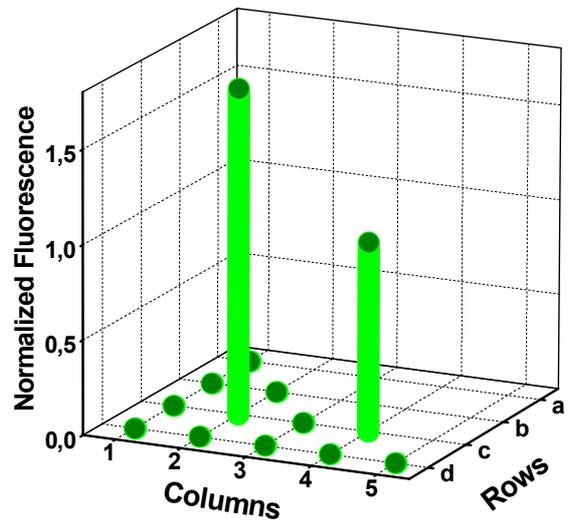

**Fig. 3. The visualisation of on-chip PCR results with fluorescent dyes.**
(A) Fluorescent image of microarray after PCR of HSV-1 and phage lambda DNAs performed in the presence of Cy-5-labelled dUTPs, taken after electrophoretical washing and subsequent drying. Bright spots in the corners contain Cy5-labelled immobilized marker oligonucleotide. (B) Fluorescent image of microarray after PCR with the same targets in presence of SYBR green I[®] taken at 72°C. (C) Quantification of data from panel A. Marker spots were not quantified. (D) Quantification of data from panel B.

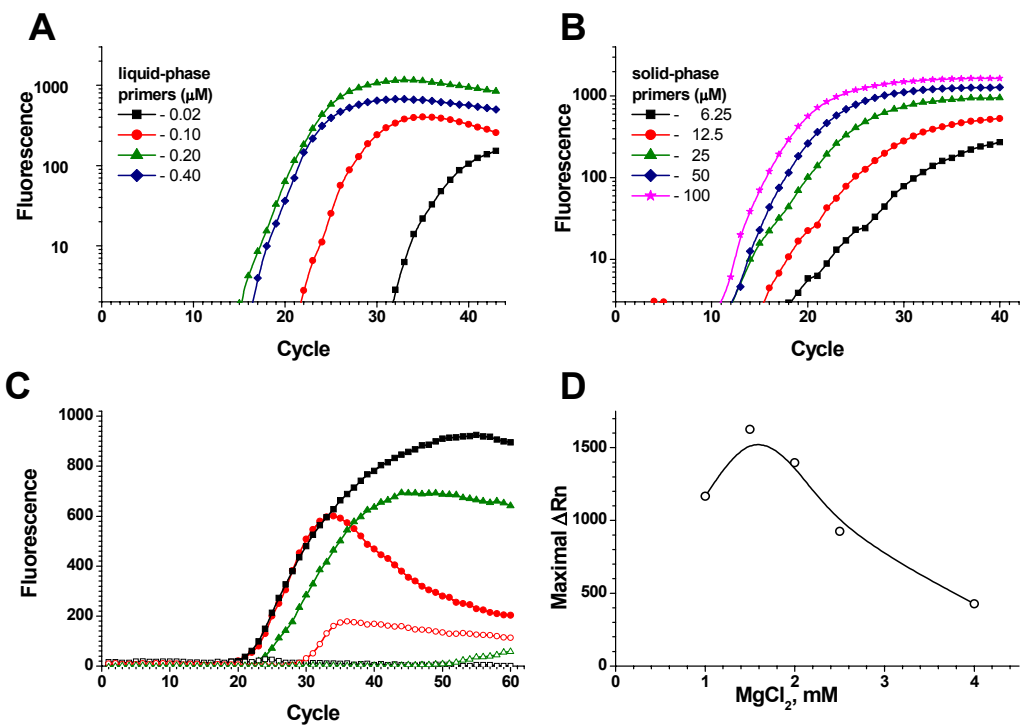

**Fig. 4. Effects of primers and magnesium concentrations on on-chip PCR characteristics.**
(A) The effect of liquid-phase primers concentrations on one-stage PCR. Liquid phase contains BL-R and HRP-R primers (Table 1) in equal concentrations indicated on the plot. (B) The effect of solid-phase primer concentration on second stage of two-stage PCR in the presence of 0.2 µM R_u primer (Table 1). (C) The effect of number of different primers in liquid phase on one-stage PCR. Closed symbols depict the fluorescence in specific pad; opened same-shaped symbols show corresponding background fluorescence measured on pads without any immobilized primers: squares – the PCR with one liquid-phase primer (BL-R); circles - the PCR in the presence of two liquid-phase primers (BL-R and HRP-R); triangles – PCR with the two liquid-phase primers in the presence of 3× PCR enhancer solution. (D) The effect of magnesium concentration on one-stage PCR. Solution phase contained one 0.2 µM BL-R primer. Lambda DNA was used as a target for all of the experiments: $6\times10^7$ copies per reaction for A, C, and D; $6\times10^2$ copies was the initial concentration in the case of two-stage reaction (panel B).

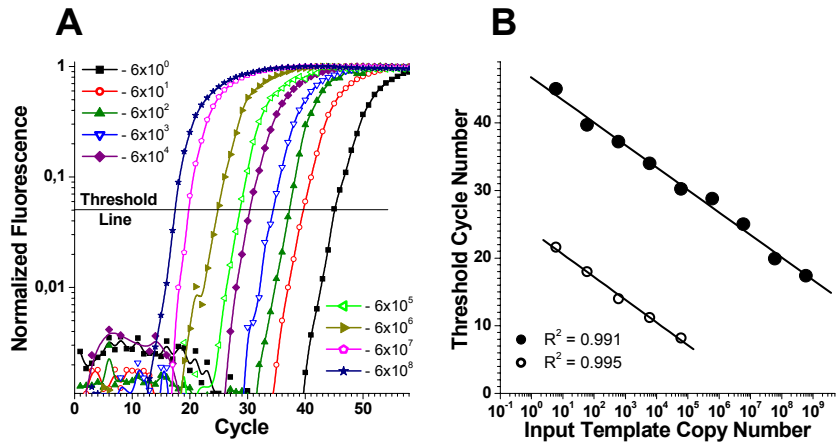

**Fig. 5. Quantitative on-chip PCR.**
(A) The amplification curves obtained for a series of 10-fold dilutions ($6 \div 6\times10^8$ copies per reaction, from left to right) of lambda DNA visualised with SYBR green I® fluorescence during one-stage on-chip PCR. (B) The calibration curves for DNA quantification. Closed circles – the calibration curve for one-stage on-chip reaction generated with $C_T$ values calculated from PCR curves of Fig. A. Opened circles – the calibration curve for second-stage on-chip PCR after 35 amplification cycles of first-stage in-tube PCR with $6 \div 6\times10^4$ initially input copies of lambda DNA.

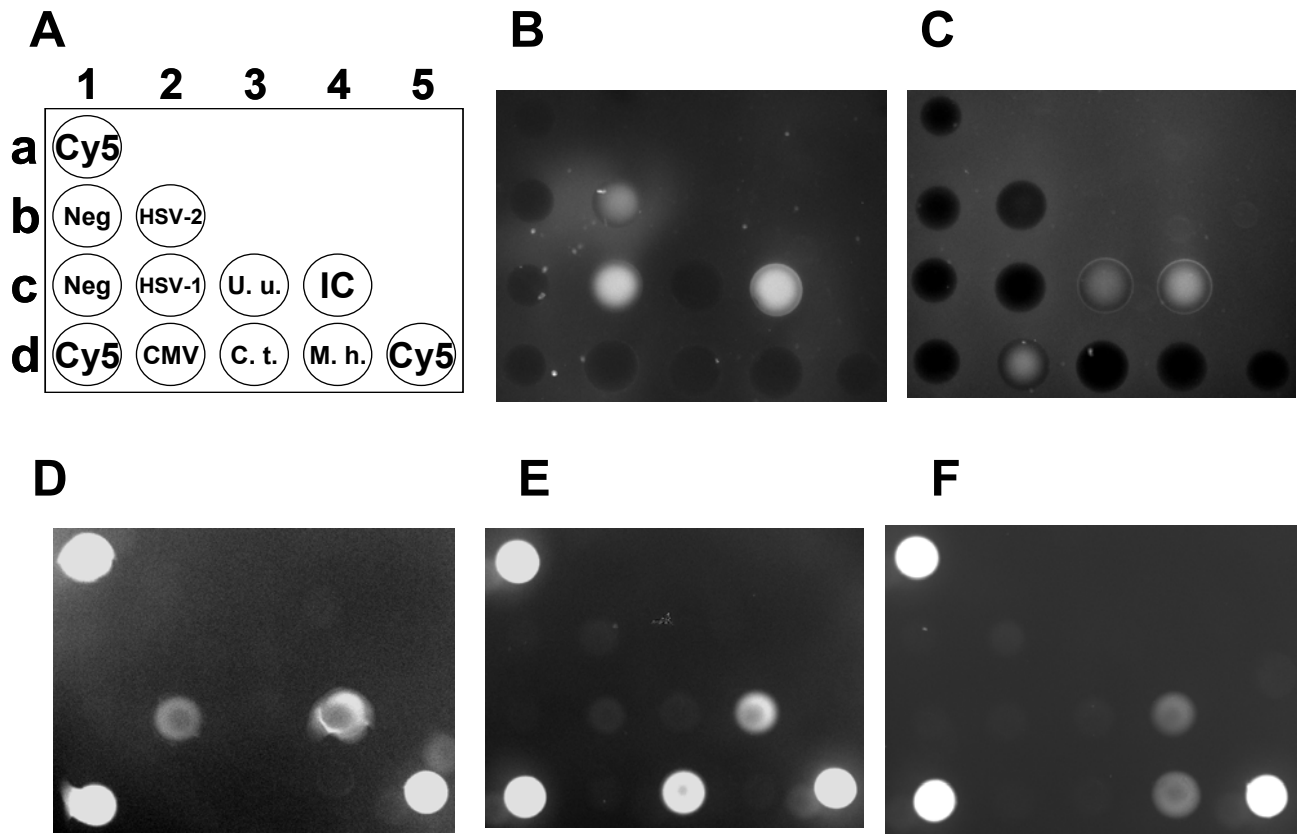

**Fig. 6. Individual results of multiplex on-chip PCR.**
(A) The template of an oligonucleotide microarray pads with solid-phase primers for identification of herpes simplex viruses types 1 and 2 ('HSV-1' and 'HSV-2', respectively), cytomegalovirus ('CMV'), *C. trachomatis* ('C. t.'), *M. hominis* ('M. h.'), *U. urealyticum* ('U. u.'), and bacteriophage lambda as an internal control ('IC'). The reference pads: 'Neg' – empty gel pads without immobilized primers; 'Cy5' – pads with 3′-immobilized Cy5-labelled non-specific oligonucleotides. (B, C) The detection of PCR results with SYBR green I®, one-stage (HSV-1 and HSV-2 positive) and second-stage PCR (CMV and *U. urealyticum* positive), respectively. (D-F) The detection of PCR results with Cy-5 labeled dUTP: D – one-stage reaction (HSV-2 positive), E, F – second-stage reactions (*C. trachomatis* and *M. hominis* positive, respectively).

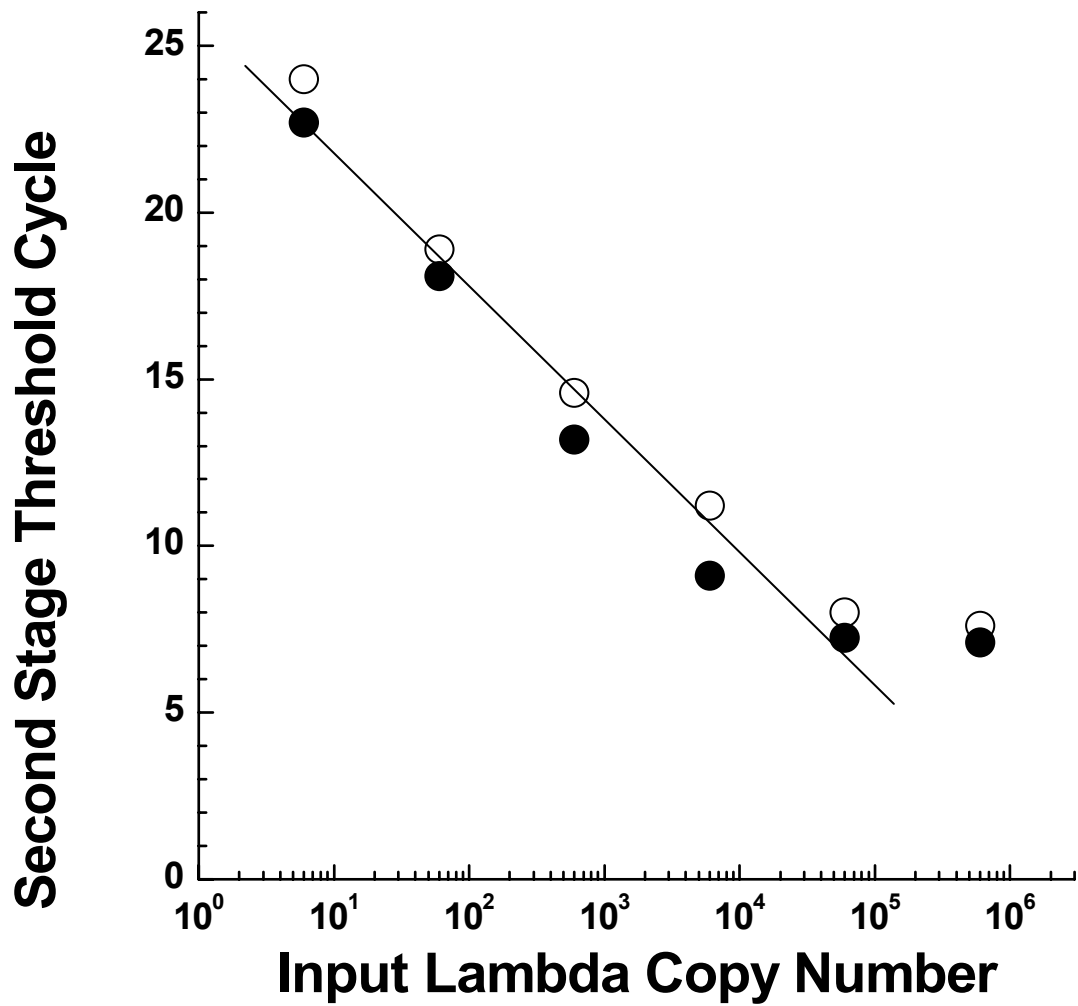

**Fig. 7. The effect of simultaneous amplification of lambda and HSV-2 DNAs on characteristics of second-stage calibration curve generated for lambda DNA quantification.**
The $C_T$ values were calculated for the second-stage on-chip PCR after 35 cycles of first-stage in-tube amplification. Closed circles – data for individual PCR of lambda DNA (6 - 6×10$^5$ copies per reaction); opened circles – PCR of lambda DNA with simultaneous amplification of abundant (about 10$^6$ copies) HSV-2 DNA.

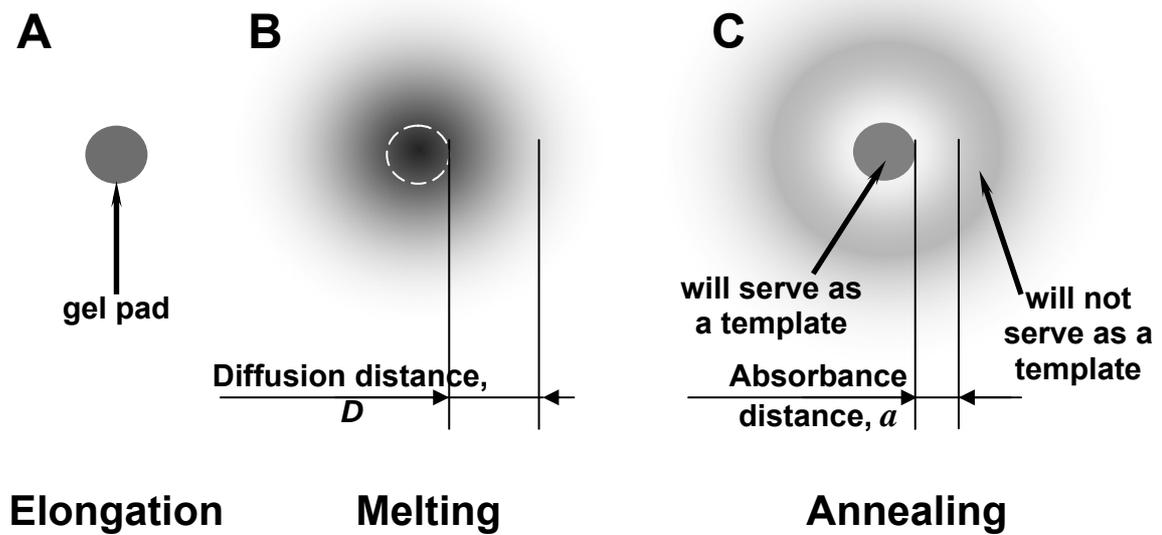

**Fig. 8. Top view of PCR product diffusion into and out of gel pad at different stages of PCR cycle.** The concentration of PCR product is given in gradation of grey. (A) Elongation stage. Newly synthesized PCR product is located in the gel pad. (B) Melting stage. PCR product diffuses out of gel pad. (C) Annealing stage. Some portion of PCR product is absorbed by gel pad and can serve as a template for the next PCR cycle. The rest of PCR product cannot participate in the next cycle of reaction because it is located out of gel pad.